\documentstyle[preprint,aps,floats,epsf]{revtex}

\tighten

\begin{document}

\draft

\title {
Backward spin polarizability $\gamma_\pi$ 
of the proton
}

\author
{
M.~Camen$^{1}$, K.~Kossert$^{1}$, F.~Wissmann$^{1}$,
J.~Ahrens$^2$, H.-J.~Arends$^2$, R.~Beck$^2$,
G.~Caselotti$^2$, P.~Grabmayr$^3$, P.D.~Harty$^4$, O.~Jahn$^2$,
P.~Jennewein$^2$, R.~Kondratiev$^5$,  M.I.~Levchuk$^6$, V.~Lisin$^5$, A.I.~L'vov$^7$, 
J.C.~McGeorge$^4$, A.~Natter$^3$, V.~Olmos de Le\'on$^2$,
M.~Schumacher$^{1}\!\!$
{\footnote{Email address: mschuma3@gwdg.de}}
, B.~Seitz$^1$, F.~Smend$^1$, A.~Thomas$^2$, 
W.~Weihofen$^1$ and F.~Zapadtka$^1$ 
}

\address
{ 
$^1$Zweites Physikalisches Institut, Universit\"at G\"ottingen, 
 D-37073 G\"ottingen, Germany
}

\address
{ 
$^2$Institut f\"ur Kernphysik, Universit\"at Mainz, 
D-55099 Mainz, Germany
}
\address
{
$^3$Physikalisches Institut, Universit\"at T\"ubingen, D-72076
  T\"ubingen, Germany
}
\address
{
$^4$Department of Physics and Astronomy, University of Glasgow,
Glasgow G12 8QQ, UK
}
\address
{
$^5$Institute for Nuclear Research, 117312 Moscow, Russia
}
\address
{
$^6$B.I. Stepanov Institute of Physics, Belarussian Academy of
Sciences, 220072 Minsk, Belarus
}
\address
{
$^7$P.N. Lebedev Physical Institute, 117924 Moscow, Russia
}

\date{\today}

\maketitle

\begin{abstract}
Using the Mainz $48{\rm cm}\, \O \!\times 64 {\rm cm}$ NaI(Tl) detector
and the segmented G\"ottingen recoil detector SENECA in coincidence, Compton
scattering by the proton at 
$\theta^{\rm lab}_\gamma = 136^\circ$ has been measured at MAMI (Mainz)
in the energy range from 200 to 470 MeV. The new data confirm the previous
observation that there is a systematic discrepancy between 
MAMI  and LEGS (Brookhaven) data leading to  
different   spin polarizabilities $\gamma_\pi =
- 38.7 \pm 1.8 $ and $-  27.2 \pm 3.1$ ($\times 10^{-4} {\rm fm}^4$), 
respectively.
\end{abstract}


The electromagnetic structure of the nucleon has been a fascinating field of
research for a long time. Great progress has been made recently
by determining basic electromagnetic structure constants with increasing
precision -- 
the electric and magnetic dipole polarizabilities $\alpha$ and $\beta$,
the forward and backward spin polarizabilities $\gamma_0$ and $\gamma_\pi$, 
and the multipole ratio E2/M1 of the radiative
$\gamma N \to \Delta$ transition.
While there is now a consistent set of data for the electromagnetic 
polarizabilities of the proton, a controversy remains for 
its  backward spin polarizability $\gamma_\pi$. 
The sum of the electromagnetic polarizabilities has been determined 
via the Baldin sum rule.
A recent redetermination of this quantity \cite{levchuk00} gave
the value
\begin{equation}
\alpha_p + \beta_p = 14.0 \pm 0.3,
\label{baldin}
\end{equation} 
(in units of $10^{-4}$fm$^3$ used for the dipole polarizabilities in the
following), which 
seems to be the most reliable result replacing previous determinations
as discussed in \cite{levchuk00}. This value is in agreement with another
recent determination \cite{olmos01} giving 
$\alpha_p + \beta_p = 13.8 \pm 0.4$.
A new value for the difference of the electric and
magnetic polarizabilities has been obtained 
by Olmos et al. \cite{olmos01} at MAMI, which, combined with previous
results, gives a  global average of
\begin{equation}
\alpha_p - \beta_p = 10.5 \pm 0.9_{\rm stat + syst} \pm 
0.7_{\rm model}.
\label{global}
\end{equation}
A detailed comparison of this number with the previous results 
may be found in  \cite{olmos01}.

A  remarkable result obtained by the LEGS group 
\cite{tonnison98,blanpied01} was the value of 
the backward spin polarizability of the proton 
$\gamma_\pi = - 27.23 \pm 2.27_{\rm stat + 
syst}{}^{+2.24}_{-2.10}{}_{\rm model}$ 
(in units of $10^{-4} {\rm fm}^4$ used for $\gamma_\pi$ throughout this
paper). This value is in contradiction with the 
predictions of standard  dispersion theory 
\cite{lvov99,drechsel98,babusci98}
and also with chiral perturbation theory
\cite{hemmert98,gellas00,kumar00} ranging from $-34$ to $-41$
 and, therefore, led to the speculation that some hitherto unknown
effect related to the spin structure of the nucleon might exist.
However, a Compton scattering experiment carried out at MAMI 
using the $48{\rm cm} \O \!\!\times 64{\rm cm}$ NaI(Tl) detector 
\cite{wissmann99} confirmed the standard value of 
$\gamma_\pi = - 37.6$ \cite{lvov97}. This latter result has 
recently been confirmed
in further experiments carried out at MAMI  using the large acceptance
arrangement LARA \cite{galler01,wolf01} and TAPS \cite{olmos01}.
The result obtained with LARA through
fits to the experimental differential cross section \cite{galler01,wolf01} is
\begin{eqnarray}
\gamma_\pi &=& -37.1 \pm 0.6_{\rm stat+syst} \pm 3.0_{\rm model},
\label{said}
\end{eqnarray}
when using the SAID-SM99K parameterization \cite{arndt96} as a basis in the
unsubstracted dispersion theory \cite{lvov97} and   
\begin{eqnarray}
\gamma_\pi &=& -40.9 \pm 0.6_{\rm stat+syst} \pm 2.2_{\rm model},
\label{maid}
\end{eqnarray}
when using the MAID2000  parameterization \cite{drechsel99}. The result obtained from low energy
Compton scattering using the TAPS detector \cite{olmos01} is
\begin{eqnarray}
\gamma_\pi &=& -35.9 \pm 2.3_{\rm stat+syst} .
\label{olmos}
\end{eqnarray}

Although there is a slight
dependence of the results obtained on the
type of parameterization of photomeson amplitudes the accuracy is good
enough to clearly not agree with the corresponding values obtained by the
LEGS group. 

It has been shown \cite{wissmann99,galler01,wolf01}
that these discrepancies between MAMI  and 
LEGS   can be traced back to a discrepancy in the experimental
differential cross sections for Compton scattering by the proton 
obtained in the $\Delta$ resonance region. In the present work
we use new data on Compton scattering by the proton,
which were obtained in the course of a series of systematic studies 
of free and quasi-free Compton scattering by the nucleon, carried out to 
determine the electromagnetic polarizabilities of the neutron, 
where the present experiments served as a test.

The apparatus used is shown in Fig.~1. Tagged photons produced in the
tagging facility at MAMI entered a scattering chamber,  containing
a  $4.6 {\rm cm} \O \!\!\times 16.3 {\rm cm}$ lq. hydrogen 
target in a Kapton 
target cell. The $48{\rm cm} \O\!\!\times 64{\rm cm}$ NaI(Tl) detector
was positioned at a distance of 60 cm from the target center
under a scattering angle of 
$\theta^{\rm lab}_\gamma$ = 136$^\circ$. As a recoil detector the G\"ottingen
SENECA detector was used, positioned at a distance of 250 cm. 
SENECA was built as a neutron detector capable
of pulse-shape discrimination. It is a honeycomb structure of 30 
hexagon-shaped detector cells of 15.0 cm minimum diameter and 20.0 cm length
filled with NE213 liquid scintillator.  The entrance face is covered
by four plastic scintillators to discriminate between charged and neutral
particles. For Compton scattering by the free proton 
carried out in the $\Delta$ range
the energy resolution
of the NaI(Tl) detector is already sufficient to discriminate between 
photons from Compton scattering and $\pi^0$ photoproduction. However,
as shown previously \cite{huenger97}, there is an 
advantage in also detecting the recoil proton
for photon energies above the peak of the $\Delta$ resonance.

The result of the experiment is shown in Fig.~2, together with theoretical
predictions
and compared with the recently published results from the LEGS experiment
\cite{blanpied01,blanpied97}, and the Mainz-LARA \cite{galler01,wolf01}
and Saskatoon \cite{hallin93}  data. For the comparison of 
the different sets of data
a scattering angle of $\theta^{\rm \, c.m.}_\gamma$ = 135$^\circ$
was chosen. 
This scattering angle is close enough to the ones of the four
experiments, so that only small corrections were needed 
to compensate for the shifts in scattering angle. These small corrections 
were calculated from the nonsubtracted dispersion theory \cite{lvov97} used 
throughout this  analysis. As an input of this 
theory the MAID2000  and SAID-SM99K parameterizations
of photomeson amplitudes have been used 
to determine the imaginary parts of invariant amplitudes.
As obtained  previously \cite{galler01,wolf01} a  mass parameter 
of ${\rm m}_\sigma$ = 600 MeV was applied 
for the pole term used to
model the asymptotic part of the invariant amplitude $A_1(\nu,t)$. 
As a further parameter of this amplitude the difference 
$\alpha -\beta = 10.5 \pm 1.1$ (see Eq.~(\ref{global}))
was used. Then the only remaining parameter is  $\gamma_\pi$.   
The backward spin polarizability extracted from the present data is
\begin{eqnarray}
\gamma_\pi&=&-36.5 \pm 1.6_{\rm stat}\pm 0.6_{\rm syst}\pm 1.8_{\rm model}
\quad \mbox{(SAID-SM99K)},\label{newsaid}\\
\gamma_\pi&=&-39.1 \pm 1.2_{\rm stat}\pm 0.8_{\rm syst}\pm 1.5_{\rm model}
\quad \mbox{(MAID2000)}    .\label{newmaid}
\end{eqnarray}
From the Mainz values of Eqs. (\ref{said}) --
 (\ref{newmaid})  given above we obtain a 
weighted average of $\gamma_\pi = - 38.7 \pm 1.8$.  
As in previous investigations
\cite{galler01,wolf01} our present values are consistent with the standard 
$\pi^0$ pole to describe  the asymptotic part of the amplitude 
$A_2(\nu,t)$ and are also in agreement with model predictions. 

{\small
This work was supported by Deutsche Forschungsgemeinschaft under
contracts SFB 201, SFB 433, Schwerpunktprogramm 1034 through contracts Wi1198, Schu222,
436 RUS 113/510.
}

\clearpage 
\newpage
\begin{figure}
\epsfxsize=0.7\textwidth
\centerline{\epsfbox{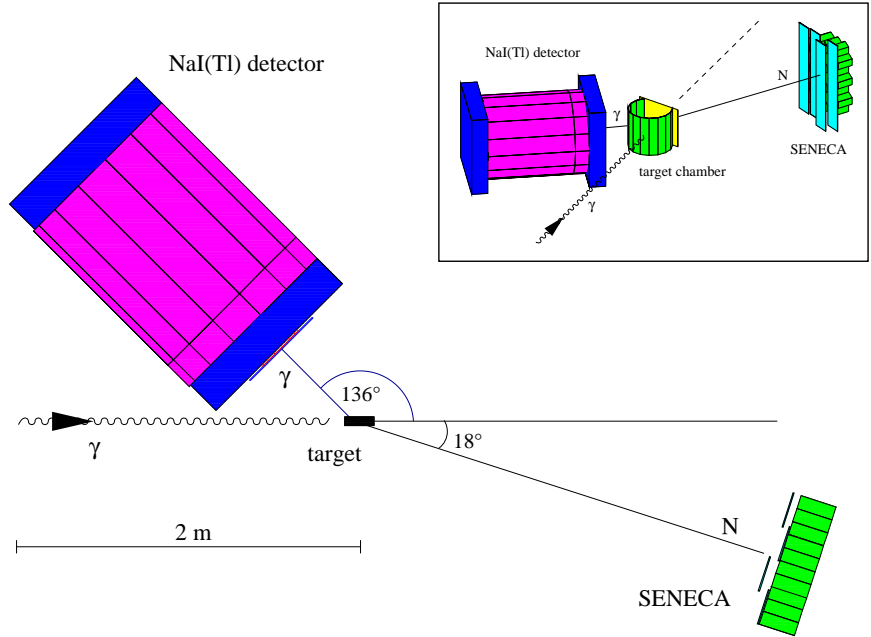}}
\caption{Experimental arrangement used for the present experiment on
Compton scattering by the proton. Compton scattering events were
identified through coincidences between the  Mainz 
48 cm $\O$ $\times$ 64 cm NaI(Tl) photon detector positioned under 
136$^\circ$ and the G\"ottingen segmented recoil counter SENECA
positioned under 18$^\circ$. 
The insert shows a perspective view of this arrangement.}
\end{figure}
\begin{figure}
\epsfxsize=0.7\textwidth
\centerline{\epsfbox{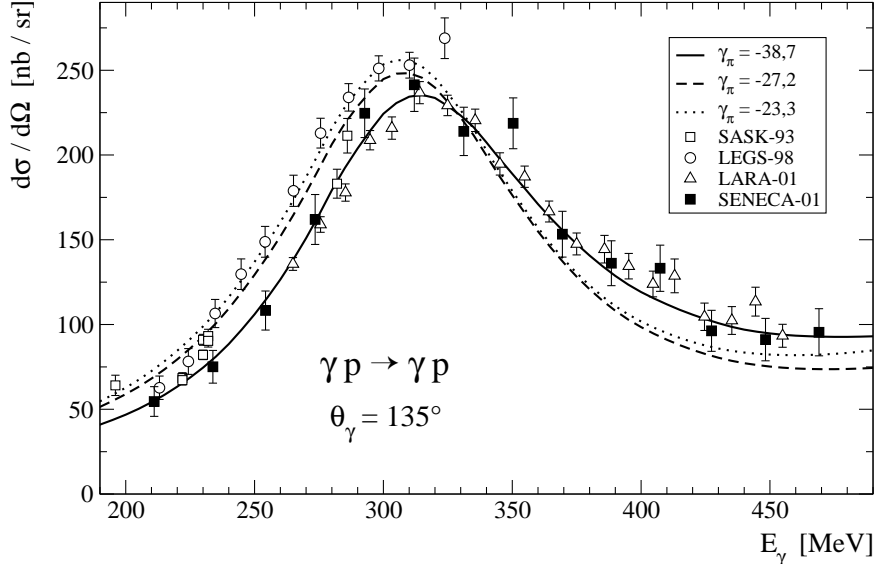}}
\caption{Experimental differential cross sections for Compton 
scattering by the proton
at $\theta^{\rm \, c.m.}_\gamma$ = 135$^\circ$. Full squares: 
present data. Open circles: LEGS data
\protect\cite{blanpied01,blanpied97}. Open triangles: Mainz-LARA data
\protect\cite{galler01,wolf01}. Open squares: Sakatoon data 
\protect\cite{hallin93}. 
The solid curve was calculated using
the nonsubtracted dispersion theory based on the 
MAID2000-parameterization  and 
adopting the Mainz weighted average of 
$\gamma_\pi = -  38.7$.
The dashed  (dotted)  curve  was obtained in the same way, 
but replacing the value 
of the spin polarizability by $\gamma_\pi = -  27.2$ ($- 23.3$),
where $\gamma_\pi = - 23.3$ corresponds to an optimal fit to the LEGS
data at $\theta^{\rm \, c.m.}_\gamma$ = 135$^\circ$ in the framework 
of the MAID2000-parameterization.} 
\end{figure}

\end{document}